\documentclass[11pt]{article}

\usepackage[letterpaper,margin=1in]{geometry}
\usepackage{microtype}
\usepackage{times}
\usepackage{amsmath,amssymb}
\usepackage{booktabs}
\usepackage{longtable}
\usepackage{listings}
\usepackage{float}
\usepackage{enumitem}
\usepackage{xcolor}
\usepackage{graphicx}
\usepackage{titlesec}
\usepackage{caption}
\usepackage[numbers,square,sort&compress]{natbib}
\usepackage[hidelinks,colorlinks=true,linkcolor=black,citecolor=blue,urlcolor=blue]{hyperref}

\lstdefinestyle{caseexcerpt}{
  basicstyle=\fontencoding{T1}\ttfamily\fontsize{8.5}{10}\selectfont,
  columns=fullflexible, keepspaces=true,
  breaklines=true, breakatwhitespace=true,
  showstringspaces=false, upquote=true,
  aboveskip=3pt, belowskip=3pt,
  frame=single, rulecolor=\color{black!20},
  backgroundcolor=\color{black!2},
  xleftmargin=4pt, xrightmargin=4pt,
  framesep=4pt
}

\newcommand{\bench}{\textsc{DoGBench}}

\titleformat{\section}{\large\bfseries}{\thesection}{0.5em}{}
\titleformat{\subsection}[block]{\normalsize\bfseries}{\thesubsection}{0.5em}{}
\titleformat{\paragraph}[block]{\normalsize\bfseries}{}{0pt}{}
\titlespacing*{\paragraph}{0pt}{1.25ex plus 0.4ex minus 0.2ex}{0.35ex}
\title{\bench: Can Agents Meet Expert Standards for User-Facing Documentation?}

\author{\small
\begin{tabular*}{0.95\textwidth}{@{\extracolsep{\fill}}ccccc}
  Frances Liu\thanks{\texttt{frances@promptless.ai}} &
  Manny Silva &
  Paige Calvert &
  Ayu Adiati &
  Sarah Sanders \\
  \textsc{Promptless} &
  \textsc{Promptless / Doc Detective} &
  \textsc{Helm} &
  \textsc{Mautic} &
  \textsc{PostHog}
\end{tabular*}
}

\date{}

\begin{document}
\maketitle

\begin{abstract}
We introduce \bench\ (Documentation Generation Benchmark), to our knowledge, the first benchmark for generating and maintaining real user-facing software documentation. It asks whether an agent can produce documentation that experienced technical writers would accept in review. The benchmark contains 292 items from open source projects, including Helm, PostHog, and Mautic. Each item gives the agent a pre-change repository and a trigger, such as a code pull request or a reported documentation gap. The agent must first decide whether the documentation needs an update. For items that need one, the agent must produce an acceptable patch in one attempt. For items that do not need updates, the agent must abstain. Task-specific rubrics, validated with project maintainers, score each patch on accuracy, completeness, reader guidance, placement, and repository conventions. The composite score combines patch quality with correct abstention, and a score of 100 means an agent meets every requirement for the task. Scores should not be interpreted as a percentage of an expert's capability. We evaluated seven agents. The highest-scoring agent reached 47.3 out of 100 on the 117-item held-out split. In a separate audit of 1,267 patches, the most common failure modes were task-completion gaps (45.5\%), technical inaccuracies (36.6\%), and incomplete conceptual or reference coverage (32.5\%). Analysis of the corresponding trajectories identified three key patterns associated with these failures: (1) describing interfaces without examining how readers use them (36.0\%), (2) missing decisive evidence and filling the gaps with plausible assumptions (33.1\%), and (3) stopping after finding the first plausible documentation surface and leaving other affected pages stale (30.1\%).
\end{abstract}

\section{Introduction}

User-facing documentation is the main public description of what a software product is, what capabilities it offers, and when and how to use them. For closed-source products in particular, documentation may be the only structured source an AI agent can use to understand and operate the product. People increasingly rely on AI agents to find, choose, and use products. Documentation therefore affects whether an agent uses a product correctly and whether the agent considers the product for the user's task at all.

Producing this content requires more than translating implementation details into prose. The work demands judgment about where information belongs, what readers are trying to accomplish, and what they already know. Teams increasingly use agents to write documentation, but no one has yet systematically evaluated the user-facing documentation that agents produce.

Prior work evaluates code-facing documentation rather than user-facing documentation (Section~\ref{sec:rw-docs}). That work covers function-level docstrings, repository-level code summaries, and internal developer documentation, and the work assumes that a human already decided that the documentation needs an update. No existing documentation benchmark tests whether an agent can tell when to leave the documentation alone.

We introduce \bench, a benchmark of 292 items from open source projects. Each item gives an agent a pre-change repository and a trigger, which is either a pull request or a reported documentation gap. The agent first decides whether the trigger calls for a documentation change. For 205 items, the correct response is a patch. For the other 87, the correct response is to abstain. Task-specific rubrics, validated with project maintainers, score each patch. The rubrics do not reward similarity to the documentation that humans merged. Depending on the task, a correct patch may revise existing guidance, create and register a new page, move content, or remove stale or redundant content. Every item starts from an existing product and its then-current documentation. The benchmark therefore does not cover writing a product's documentation from scratch or redesigning an information architecture without constraints.

We use a random, stratified 117-item held-out split for primary evaluation. The other 175 items form a public development split. We use the public split and the full 292 items only for robustness analyses.

We evaluated seven agent lanes. The highest-scoring agent reached 47.3 out of 100 on the held-out split. Only 6.1\% of submissions contained fabricated content. The more common failure was a plausible patch that left the reader unable to finish the task. Of 1,267 submissions, 45.5\% had a task-completion gap. Section~\ref{sec:failure-analysis} traces these failures to how agents investigate. In 36.0\% of submissions, agents explained product interfaces without checking how readers use them. In 33.1\%, agents stopped before finding decisive evidence. In 30.1\%, agents edited the first plausible documentation surface and missed other surfaces the change affected.

\section{Related Work}

\subsection{Documentation generation benchmarks}
\label{sec:rw-docs}
Existing documentation benchmarks focus on code-facing documentation. CodeSearchNet supplied a corpus of paired functions and documentation for semantic code search, and CodeXGLUE used CodeSearchNet-derived data for code summarization~\citep{husain2019codesearchnet,lu2021codexglue}. More recent benchmarks study docstring updates after code changes (CoDocBench) and repository-level internal documentation (CodeWikiBench)~\citep{pai2025codocbench,nguyen2025codewikibench}. Like \bench, SWD-Bench builds its tasks from pull requests, but it scores repository-level documentation by how well a model can use that documentation to answer questions about the repository's functionality~\citep{wang2026swdbench}. None of these benchmarks asks the model to decide whether documentation needs an update.

\subsection{Scoring open-ended edits}
SWE-bench, which also draws from open source repositories, is widely used to evaluate code generation~\citep{jimenez2024swebench}. OpenAI has since questioned the validity of its Verified subset because narrow tests can reject correct alternative solutions~\citep{openai2026swebenchverified}. That risk is larger for documentation because many different edits can satisfy the same reader need. \bench\ therefore scores each patch against requirements drawn from the triggering change and the pre-change repository rather than against similarity to the merged human patch.

\section{Benchmark Design}

\begin{figure}[t]
  \centering
  \includegraphics[width=\textwidth]{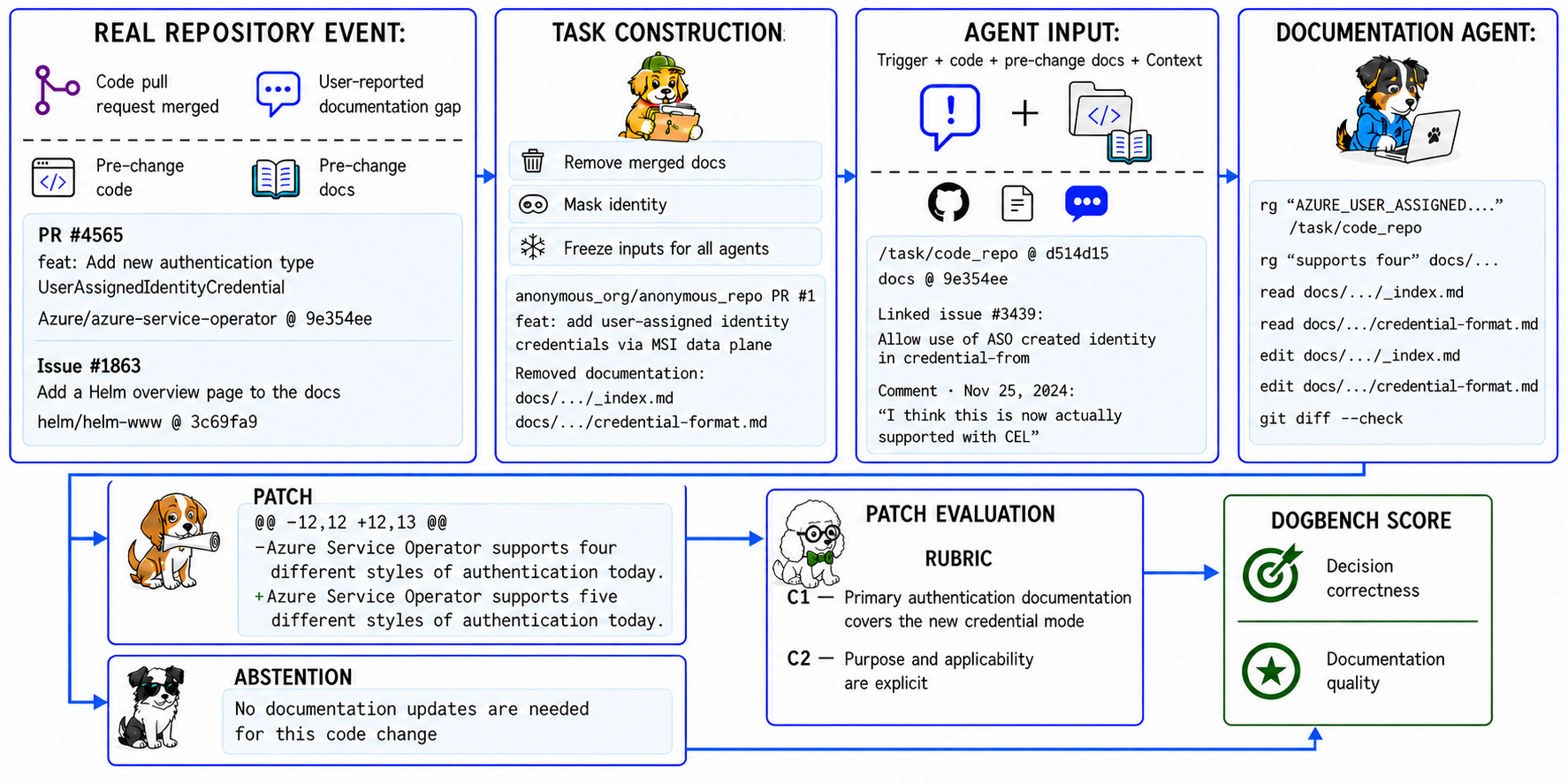}
  \caption{Overview of \bench. Each item begins with a real trigger event and a frozen, identity-masked snapshot of the pre-change repository. A documentation agent either abstains or emits a patch, and task-specific rubrics score the patch. \bench\ reports decision correctness and documentation quality separately. Its composite score combines delivered patch quality with abstention recall.}
  \label{fig:dogbench-overview}
\end{figure}

Each item in the benchmark gives the agent a pre-change repository and a trigger. The trigger is either a code pull request or a user-reported documentation gap, often a GitHub issue. This setup mirrors how maintainers work. A maintainer either ships documentation changes with a feature or updates the documentation in response to a community issue. The agent must either return a patch that edits the documentation or abstain from making documentation changes when no user-facing change is needed. Figure~\ref{fig:dogbench-overview} summarizes the item-construction and evaluation pipeline.

\subsection{Dataset construction}
\label{sec:dataset}
The benchmark contains 292 items: 205 require a documentation change, and 87 require abstention. We built them from three pools. The first holds 90 changes initially sampled as likely abstention cases. The second holds 136 code-triggered documentation updates, and the third holds 66 explicit user-reported documentation gaps. During final adjudication, we reclassified three of the 90 likely abstention cases as requiring documentation, which left 87 abstention items. The final set therefore has 139 code-triggered documentation items, 66 items triggered by reported documentation gaps, and 87 abstention items. Project maintainers reviewed 42 items from repositories such as Helm, Doc Detective, Mautic, and PostHog. Appendix~\ref{app:dataset-details} gives additional selection details and two case studies.

\paragraph{Constructing the trigger}
For a pull request that ships code and documentation together, we remove the documentation changes and give the agent only the code change. For a pull request that changes only documentation, we build the trigger from linked issues and discussions. We add sanitized versions of the source pull request's title and description. Both methods keep the documentation need and the reason for it but hide how the maintainer wrote the documentation.

We selected open source repositories with English-language documentation across diverse ecosystems. We exclude the following:
\begin{itemize}[nosep]
    \item reverts, release-only version bumps, and merge or sync pull requests
    \item pure refactors
    \item changes where the connection between trigger and the documentation need is unclear
    \item items that need context unavailable in the public repository or trigger
\end{itemize}
We also exclude bot-authored pull requests are excluded unless a human maintainer reviewed and revised the change before merging. Appendix~\ref{app:dataset-details} gives the full rules and known limitations.

\subsection{Contamination controls}
\label{sec:contamination}
Because the source events are public, contamination can happen if a model has seen the merged documentation during training or if an agent finds it while running.  We probe training-time exposure with source-event-date and repository-footprint ablations (Section~\ref{sec:results-ablations})~\citep{sainz2023contamination}. To prevent execution-time exposure, agents work in fresh Docker containers with identity-masked repositories and no network access except to the model provider. We also inspect agent trajectories for attempts to reach the merged human patch. A conservative overlap detector flags suspicious similarity to the merged documentation for manual review. We found no confirmed case of copying. We reviewed every retained detector alert and judged each one a false positive.

\section{Evaluation Protocol}
\label{sec:scoring-model}

For each item that requires a documentation update, we score the submitted patch against a task-specific rubric. Across the 205 documentation-needed items, the rubrics contain 3,273 criteria, including 798 P0 criteria.

We design each criterion to test one observable review decision. For example, suppose a change lets a Helm values file to contain multiple YAML documents. Separate criteria can then check the documentation for three statements: documents are processed in order, later values take precedence, and nested maps are merged recursively. A broad criterion such as ``explains multi-document values well'' would not be testable enough.

\subsection{Criterion types and priorities}

Each criterion use one of three scoring types:
\begin{itemize}[leftmargin=1.6em,itemsep=2pt,topsep=3pt,parsep=0pt,partopsep=0pt]
\item A \textbf{\emph{requirement}} always applies and receives a binary Pass or Fail verdict.
\item A \textbf{\emph{conditional criterion}} applies only when the patch meets its condition. We call a conditional criterion \emph{triggered} when it applies.
\item A \textbf{\emph{deduction-only guardrail}} prohibits content such as a fabricated command or an unsafe recovery step. Avoiding the prohibited content earns no credit, and introducing it costs a deduction.
\end{itemize}
Requirements and triggered conditional criteria receive only Pass or Fail verdicts, with no partial credit. Untriggered conditional criteria are left out of scoring.

Each criterion also has a priority from P0 to P3, which is separate from its scoring type. P0 is reserved for a defect that blocks the patch on its own: failing that criterion alone would require revision under the benchmark's standard. Typical P0 defects include the following:
\begin{itemize}[leftmargin=1.6em,itemsep=2pt,topsep=3pt,parsep=0pt,partopsep=0pt]
\item materially misstating product behavior
\item omitting information the reader needs to complete the central reader task
\item giving an unsafe or destructive instruction
\item inventing a public interface
\item leaving an essential maintained documentation surface contradictory or unusable
\end{itemize}

Optional examples, secondary edge cases, stylistic preferences, and exact wording do not qualify as P0 merely because they would improve the patch. A patch is \emph{P0-clean} when no applicable P0 criterion failed. P0-clean diagnoses only critical defects and does not a prediction whether a maintainer would merge the patch unchanged.

\subsection{Score construction}

Let $R$ contain all requirements and triggered conditional criteria, and let $G$ contain all violated deduction-only guardrails. Each criterion in $R$ contributes one point if it passes, and each violated guardrail deducts one point. For $|R|>0$, the uncapped patch score is
\[
\widetilde q =
100\,\frac{\max\!\left(0,\sum_{i\in R}\mathbf{1}[i\text{ passes}]-|G|\right)}
{|R|}.
\]
An untriggered conditional criterion counts in neither the numerator nor the denominator. A guardrail that the patch does not violate is also left out, so a patch earns no points merely for avoiding an optional risk. If $R$ is empty, the score is zero.

Let $B=1$ when a P0 requirement or triggered conditional criterion fails, or when a P0 guardrail is violated. Otherwise, let $B=0$. The reported patch score is
\[
q =
\begin{cases}
\min(\widetilde q,60), & B=1,\\
\widetilde q, & B=0.
\end{cases}
\]
On a documentation-needed item, an empty patch or an abstention also scores zero. We set the 60-point ceiling as an evaluation policy. We did not estimate it from maintainer editing time or acceptance decisions. The ceiling prevents success on many secondary criteria from averaging away a critical defect.

\subsection{Rubric construction}
\label{sec:rubric-eval}

\paragraph{Research and synthesis}
We build each rubric through repository research and an LLM-council process. The merged human patch is not included among the candidate patches supplied to the rubric agents. A rubric research agent inspects the triggering change and the repository to identify what users need to know and the evidence that supports each requirement. The research agent has internet access and may encounter the merged human patch, but every criterion must have independent supporting evidence; the human patch alone cannot justify a criterion. A synthesis stage turns these findings into criteria with explicit passing and failing conditions. \citet{mei2026drrubric} also explore this research-to-criteria approach.

\paragraph{Differential review}
The differential stage compares anonymous candidate patches side by side to find editorial choices that the draft rubric does not yet cover. This stage follows the observation that inspecting model outputs can help refine evaluation criteria~\citep{shankar2024validators}. For each uncovered difference, the rubric agent checks the research findings and gathers more evidence where needed. It then decides whether the difference matters to the reader. A difference between candidates only raises a question and does not establish what is correct. Trivial or neutral differences do not become criteria. The rubric agent also records supported documentation needs that no candidate meets. These findings are then used to revise the draft rubric.

\paragraph{Audit and debate}
A second model, from a different model family, audits the revised criteria and their priorities. When the rubric author and the auditor disagree, they revisit the evidence and exchange arguments. Together they revise or remove any criterion that cannot be justified. The debate ends when the auditor accepts the revised rubric or the exchange reaches its configured limit. The longest saved debate we inspected ran 28 messages after the opening audit.

\subsection{Human validation}

We compare the resulting rubrics with independently collected maintainer criteria on a reviewed subset. Separately, we compare the Pass or Fail verdicts of a scoring model with human judgments. The first check asks whether the rubric captures the requirements that maintainers consider important. The second asks whether the scoring model applies those requirements correctly.

On 42 maintainer-reviewed items, the automatic documentation-need gate agrees with the maintainers on 41, with one false positive. On 21 documentation-needed items with completed criterion alignment, mean priority-weighted recall against maintainer criteria is 0.892. In a separate study of 20 items and 330 criteria, the scoring model agrees with the post-adjudication human reference on 310 criteria (93.9\%). One paper author resolved disagreements in the human reference after review, so this figure is not blinded agreement between two humans. Appendix~\ref{app:evaluation-details} describes both validation studies.

\section{Experimental Setup}
\label{sec:baselines}

We evaluate seven agent lanes:
\begin{itemize}[leftmargin=1.6em,itemsep=2pt,topsep=3pt,parsep=0pt,partopsep=0pt]
\item GLM 5.2, Qwen3.8 Max, and Kimi K2.7 Code with OpenCode
\item Claude Opus 4.8 and Claude Sonnet 4.6 with Claude Code
\item GPT-5.5 and GPT-5.6 Sol with Codex
\end{itemize}
Every agent runs each item in a fresh Docker environment with the same inputs. The inputs are the pre-change code and documentation repository, plus the trigger: a code diff or a reported documentation gap. Each agent--item pair runs once and must either return a patch or abstain. Agents can use command-line tools to inspect and edit the repository, but the environment has no network access except to the model providers.

\paragraph{Why the environment is sealed}
Internet access can be valuable for documentation agents in ordinary use, but \bench\ blocks it to prevent contamination. Because the source events and merged documentation are public, a connected agent could retrieve the answer instead of solving the task from the supplied evidence. In an audit of an earlier version of the benchmark, we found that 31\% of runs retrieved the source pull request despite identity masking. A further 6\% copied directly from other agents' earlier trajectories because the agents shared a host.

\paragraph{Data splits}
The release has two splits. The development split holds 175 items: 123 documentation-needed and 52 abstention. The held-out split holds 117 items: 82 documentation-needed and 35 abstention. The random split draw comes from a procedure stratified by class, source pool, task type, and repository. We did not choose the split by inspecting measured scores. Development items include inputs, rubrics, and reference artifacts. Held-out rubrics, references, and item-level scores remain private. We report only the seven reproducible agents in this paper.

\section{Results}

Table~\ref{tab:headline-composite} reports the composite score on the 117-item held-out split. The composite score combines two capabilities: delivering useful patches when documentation is needed and correctly refraining from editing when it is not. The table reports the following measures:

\begin{description}[style=nextline,leftmargin=1.5em,labelindent=0pt,itemsep=3pt,topsep=3pt]
\item[Delivered quality.] Delivered patch quality, $D$, averages rubric scores across the 82 held-out items that require an update. A missed or empty patch scores zero. Conditional quality averages only valid emitted patches.
\item[Abstention recall.] Abstention recall, $N$, measures correct abstention across the 35 held-out items that need no update.
\item[Composite score.] We combine $D$ and $N$ with the harmonic mean, $C=2DN/(D+N)$. The harmonic mean treats useful patches and correct abstention as jointly necessary. It also keeps the score independent of the benchmark's constructed class proportions.
\item[Accuracy.] Decision accuracy covers all 117 items.
\item[P0-clean delivery.] P0-clean delivery is the share of the 82 documentation-needed items that received a valid P0-clean patch (Section~\ref{sec:scoring-model}).
\end{description}

Qwen3.8 Max+OpenCode has the highest composite score (47.3), followed by GPT-5.6 Sol+Codex (46.2) and GLM 5.2+OpenCode (44.2). GPT-5.6 Sol+Codex has the highest P0-clean delivery (39.0). Appendix~\ref{app:analysis-details} reports results on all 292 items separately.

\begin{table}[H]
\centering
\footnotesize
\setlength{\tabcolsep}{3.2pt}
\renewcommand{\arraystretch}{1.15}
\begin{tabular}{@{}lrrrrrrr@{}}
\toprule
\textbf{Agent} & \textbf{Score} & \textbf{Accuracy} &
\shortstack{\textbf{Patch}\\\textbf{recall}} &
\shortstack{\textbf{Abstention}\\\textbf{recall}} &
\shortstack{\textbf{P0-clean}\\\textbf{delivery}} &
\shortstack{\textbf{Delivered}\\\textbf{quality}} &
\shortstack{\textbf{Conditional}\\\textbf{quality}} \\
\midrule
\shortstack[l]{Qwen3.8 Max\\+OpenCode} & 47.3 & 79.5 & 80.5 & 77.1 & 26.8 & 34.1 & 42.3 \\
\shortstack[l]{GPT-5.6 Sol\\+Codex} & 46.2 & 81.2 & 95.1 & 48.6 & 39.0 & 44.1 & 46.3 \\
\shortstack[l]{GLM 5.2\\+OpenCode} & 44.2 & 83.8 & 84.1 & 82.9 & 23.2 & 30.1 & 35.8 \\
\shortstack[l]{Kimi K2.7 Code\\+OpenCode} & 43.9 & 82.1 & 91.5 & 60.0 & 24.4 & 34.6 & 37.8 \\
\shortstack[l]{Claude Opus 4.8\\+Claude Code} & 41.2 & 81.2 & 79.3 & 85.7 & 24.4 & 27.1 & 34.2 \\
\shortstack[l]{Claude Sonnet 4.6\\+Claude Code} & 40.0 & 82.9 & 93.9 & 57.1 & 20.7 & 30.8 & 32.8 \\
\shortstack[l]{GPT-5.5\\+Codex} & 34.0 & 76.1 & 96.3 & 28.6 & 36.6 & 42.0 & 43.6 \\
\bottomrule
\end{tabular}
\captionsetup{font=footnotesize}
\caption{Primary results on the 117-item held-out split (\%). Rows are ordered by the unrounded composite score.}
\label{tab:headline-composite}
\end{table}

\subsection{Patch or abstention decision performance}
\label{sec:results-decision}

The held-out decision task contains 82 documentation-needed items (70.1\%) and 35 abstention items (29.9\%). Table~\ref{tab:decision-performance} treats \texttt{patch} as the positive class.

\begin{table}[H]
\centering
\small
\begin{tabular}{lrrr}
\toprule
\textbf{Agent} & \textbf{Accuracy} & \textbf{Patch recall} & \textbf{Abstention recall} \\
\midrule
GLM 5.2+OpenCode & 83.8 (98/117) & 84.1 (69/82) & 82.9 (29/35) \\
Claude Sonnet 4.6+Claude Code & 82.9 (97/117) & 93.9 (77/82) & 57.1 (20/35) \\
Kimi K2.7 Code+OpenCode & 82.1 (96/117) & 91.5 (75/82) & 60.0 (21/35) \\
GPT-5.6 Sol+Codex & 81.2 (95/117) & 95.1 (78/82) & 48.6 (17/35) \\
Claude Opus 4.8+Claude Code & 81.2 (95/117) & 79.3 (65/82) & 85.7 (30/35) \\
Qwen3.8 Max+OpenCode & 79.5 (93/117) & 80.5 (66/82) & 77.1 (27/35) \\
GPT-5.5+Codex & 76.1 (89/117) & 96.3 (79/82) & 28.6 (10/35) \\
\bottomrule
\end{tabular}
\captionsetup{font=footnotesize}
\caption{Patch or abstention decisions for seven lanes on the 117-item held-out split (\%). Parentheses give the numerator and denominator. Patch recall measures recovery of required updates, and abstention recall measures correct abstention.}
\label{tab:decision-performance}
\end{table}

Abstention recall ranges from 28.6\% to 85.7\%. GLM 5.2+OpenCode has the highest held-out decision accuracy (83.8\%). GPT-5.5+Codex recovers 96.3\% of required patches, and GPT-5.6 Sol+Codex recovers 95.1\%. They make 25 and 18 incorrect decisions, respectively, on the 35 abstention items.

Unnecessary edits have real cost for both the documentation reader and the maintainers. These edits can add implementation details that users neither need nor can act on, which bloats the documentation and makes relevant guidance harder to find. Every unnecessary patch also needs maintainer attention during triage, review, and ongoing maintenance, and it adds work to downstream tasks such as translation and versioning. Abstaining from an unwarranted edit is therefore a documentation-quality and governance requirement, and we think the benchmark should measure it.

\subsection{Documentation quality results}
\label{sec:results-documentation-quality}

Table~\ref{tab:documentation-quality} reports quality conditional on a correct patch decision. GPT-5.6 Sol+Codex (46.3) and GPT-5.5+Codex (43.6) have the highest means.

\begin{table}[H]
\centering
\begin{tabular}{lrrrr}
\toprule
\textbf{Agent} & \textbf{Mean} & \textbf{95\% CI} & \textbf{Median} & \textbf{Correct patches ($n$)} \\
\midrule
GPT-5.6 Sol+Codex & 46.3 & [40.8, 51.3] & 50.0 & 78 \\
GPT-5.5+Codex & 43.6 & [37.8, 48.7] & 42.9 & 79 \\
Qwen3.8 Max+OpenCode & 42.3 & [35.0, 48.7] & 42.8 & 66 \\
Kimi K2.7 Code+OpenCode & 37.8 & [31.9, 43.3] & 35.7 & 75 \\
GLM 5.2+OpenCode & 35.8 & [29.7, 41.8] & 35.7 & 69 \\
Claude Opus 4.8+Claude Code & 34.2 & [28.5, 40.1] & 36.4 & 65 \\
Claude Sonnet 4.6+Claude Code & 32.8 & [27.7, 38.1] & 33.3 & 77 \\
\bottomrule
\end{tabular}
\captionsetup{font=footnotesize}
\caption{Scored documentation quality on the 82 documentation-needed held-out items, conditional on a correct patch decision. Empty outputs, abstentions, and wrong decisions receive no quality score here. Their cost appears in delivered patch quality and the composite score. Intervals are percentile 95\% intervals from 20,000 bootstrap resamples of documentation repositories.}
\label{tab:documentation-quality}
\end{table}

\subsection{Ablations}
\label{sec:results-ablations}

\noindent\textbf{Source-event-date ablation}\label{sec:results-mergedate}
We use source-event dates to probe training-time exposure: pull-request merge dates and issue creation dates. For each agent whose model has a provider-published knowledge cutoff, we divide all 205 documentation-needed items at the cutoff. All four agents score lower on post-cutoff items, but every repository-clustered interval includes zero, so the comparison provides no statistically conclusive evidence of a cutoff effect.

\begin{table}[H]
\centering
\small
\begin{tabular}{lrrrrrr}
\toprule
\textbf{Agent} & \textbf{Pre $n$} & \textbf{Pre score} & \textbf{Post $n$} &
\textbf{Post score} & \textbf{$\Delta$} & \textbf{Repo 95\% CI} \\
\midrule
Claude Opus 4.8+Claude Code & 86 & 29.4 & 119 & 27.1 & $-2.2$ & [$-9.5$, 4.8] \\
Claude Sonnet 4.6+Claude Code & 50 & 34.6 & 155 & 30.9 & $-3.7$ & [$-12.8$, 5.3] \\
GPT-5.5+Codex & 69 & 42.9 & 136 & 40.9 & $-2.0$ & [$-9.0$, 7.1] \\
GPT-5.6 Sol+Codex & 94 & 44.3 & 111 & 44.0 & $-0.3$ & [$-6.6$, 6.7] \\
\bottomrule
\end{tabular}
\caption{Mean documentation-quality score before and after each model's published knowledge cutoff, on all 205 documentation-needed items. Event dates are pull-request merge dates or issue creation dates. $\Delta$ is the post-cutoff score minus the pre-cutoff score, and intervals resample repositories.}
\label{tab:cutoff-ablation}
\end{table}

\noindent\textbf{Repository-footprint ablation}\label{sec:results-obscure}
We also compare 82 items from low-footprint repositories (fewer than 1,000 GitHub stars) with 91 items from popular repositories (over 5,000 GitHub stars). The comparison shows no consistent advantage for popular repositories. Table~\ref{tab:popularity-ablation} in the appendix gives per-agent values.

\noindent\textbf{Documentation-scale ablation}\label{sec:results-doc-scale}
We also tested whether larger documentation sets affect scores. Across repository-level means, doubling the page count changes the score by $-0.98$ points (95\% CI $-2.81$ to $+0.69$). The interval includes zero, so the data show no clear relationship between documentation size and score.

\section{Failure Analysis}
\label{sec:failure-analysis}

\subsection{Patch/abstention decision failures}
\label{sec:decision-failure-mechanisms}

Agents make decision failures by making edits when no change is needed (\emph{over-editing}) and by failing to edit when change is needed (\emph{under-editing}).

Over-editing often starts with a misleading cue. The cue is an internal symbol with a user-facing-sounding name, or an existing documentation page that mentions the affected component. The agent treats that cue as proof that the change alters documented user behavior. Typically, the agent documents one of the following, none of which needs documentation:

\begin{itemize}[leftmargin=1.5em,labelsep=0.5em,itemsep=1pt,topsep=3pt,parsep=0pt,partopsep=0pt]
\item an internal refactor, rename, or mechanically regenerated type whose public contract is unchanged
\item a performance optimization, test, CI, or dependency change with no observable user effect
\item generated files or reference artifact that should not be manually updated
\end{itemize}
These patches often look defensible because the terms and destination are topically related to the diff. The missing step is establishing that the change creates behavior a user can invoke, observe, or act on. Once an agent predicts that it should edit, it tends to search for somewhere to put text. It does not go back to ask whether a documentation obligation exists, even when it later finds evidence for the opposite decision.

\medskip
Under-editing happens when agents treat implementation location, the absence of existing coverage, or weak keyword overlap as evidence that no documentation change is needed. These proxies cause agents to miss user-facing changes. Common forms include the following:

\begin{itemize}[leftmargin=*,itemsep=2pt,topsep=2pt]
\item treating a real public interface, such as a SQL keyword, API function, configuration option, or credential setting, as an implementation detail because it lives in a parser, registry, or configuration module
\item reading a new data source, integration, or landing-page capability as internal pipeline plumbing rather than a change to the reader's available workflow
\item treating a failed search for existing coverage as evidence that a subject is intentionally undocumented, when the missing coverage may be the documentation gap the task exposes
\item declining documentation-only or issue-driven work because there is no implementation diff, even when the task identifies a verified gap in existing behavior
\end{itemize}

The \emph{absence-as-evidence} case reinforces itself. In incorrect-abstention trajectories, an agent searches the existing documentation for the feature, interface, or workflow that the task affects. It finds no coverage and reads that silence as evidence that the subject is intentionally out of scope. From the agent's perspective, deliberate exclusion and an unfilled gap produce the same search result. Once the agent treats absence as policy, the omission propagates.

\subsection{Documentation quality failures}
\label{sec:failure-modes}
\label{sec:frontier-failures}
\label{sec:writing-failure-mechanisms}

\begin{table}[H]
\centering
\footnotesize
\setlength{\tabcolsep}{5pt}
\renewcommand{\arraystretch}{1.08}
\begin{tabular}{p{0.25\textwidth}p{0.55\textwidth}r}
\toprule
\textbf{Technical-writing failure mode} & \textbf{Effect on the documentation} &
\textbf{Rate} \\
\midrule
Task-completion gap & Omits a decision point, procedure, verification step, or recovery path needed
to complete the reader's task. & 45.5\% \\
Technical inaccuracy & Misstates an interface or behavior, including its scope, default,
lifecycle, or compatibility. & 36.6\% \\
Incomplete conceptual or reference coverage & Omits the core concept, capability, or contract the
reader needs to understand and use the change. & 32.5\% \\
Missing supporting information & Covers the main task but omits material rationale, boundaries,
examples, operational detail, or secondary cases. & 26.5\% \\
Missing prerequisites & Omits permissions, dependencies, credentials, versions, resources,
enablement, or other setup conditions. & 21.0\% \\
Information-architecture or findability failure & Places content outside the reader's likely path
or omits navigation, cross-references, and findable terminology. & 20.7\% \\
Missing audience and purpose framing & Does not establish who the content is for, why it matters,
or when to use it. & 19.9\% \\
Cross-surface content inconsistency & Updates one surface while leaving an authoritative, mirrored,
generated, or linked surface stale. & 13.3\% \\
\bottomrule
\end{tabular}
\captionsetup{font=footnotesize}
\caption{Share of submissions with each patch-level problem, across the 1,267 submissions audited before the reruns. The table lists labels assigned to at least 10\% of submissions, and Table~\ref{tab:full-patch-failure-taxonomy} in the appendix lists the rest. One patch can have several problems. The labels describe defects in the patch itself. Table~\ref{tab:trajectory-root-causes} covers the causes in the agents' trajectories. Only 1.3\% of submissions were labeled ``no material defect'' among the selected top labels.}
\label{tab:patch-failure-modes}
\end{table}

\paragraph{Hallucinations often distort real behavior}
Our audit labeled only 6.1\% of submissions as containing fabricated content, a category that includes invented classes, flags, and endpoints. Technical inaccuracies appeared in 36.6\% of submissions and involved scope, defaults, lifecycle, or compatibility. The failure taxonomy classifies invented interfaces or capabilities as fabricated content and false descriptions of real interfaces as technical inaccuracies. In practice, the boundary is not always clear.

One agent wrote that Helm 4 uses Server-Side Apply by default when installing or upgrading releases.\footnote{Item \texttt{helm-helm-www-pr1926}, Claude Sonnet 4.6+Claude Code.} That default applies to new installations. Releases created with Helm 3 continue using client-side apply after upgrading unless the user explicitly switches them. The agent explained this distinction later in the patch, but its opening statement still gave readers the wrong default. We classify this error as a technical inaccuracy because the feature exists. Extending the feature's behavior beyond its supported conditions could also reasonably count as hallucination. In many audited cases, the agent took behavior that holds under narrow conditions and presented it as true in general. These claims are unfounded, like hallucinations, but they distort real functionality instead of inventing it.

\paragraph{Agent patches often lack a model of the reader's task}
Agents can often describe a feature's behavior. They are less able to write for a reader who came to the documentation to decide something or reach a goal. Audience and purpose framing is missing in 19.9\% of audited submissions. These patches explain what a feature does but not who should use it, why it is useful, or when to choose it. For example, one Strawberry GraphQL patch correctly documented how to select an older Apollo Federation version. It did not explain why a reader might need to: to upgrade Strawberry while staying compatible with an older Apollo Router or Gateway.\footnote{Item \texttt{strawberry-graphql-strawberry-pr4045}, GPT-5.5+Codex.} The patch documented the setting but omitted the decision it was designed to support.

This limitation extends beyond explaining when or why to use a feature. Much technical documentation guides readers through a task, and the reader's goal is to complete that task. Doing so may require prerequisites, intermediate decisions, procedural steps, verification, and recovery guidance. Task-completion gaps appear in 45.5\% of audited submissions, and missing prerequisites in 21.0\%. These failures suggest that agents treat a change as one piece of information to convey rather than as one part of a larger user journey. A patch may therefore describe the behavior accurately and still leave the reader unable to accomplish the task that brought them to the documentation.

This narrow view also affects how agents treat the documentation as a whole. Readers, both humans and agents, reach a page through search, navigation, related guides, and examples. Agents may add accurate information to a page that the intended reader is unlikely to visit, create a page without linking it from the relevant workflow, or update one surface and leave another surface on the same topic stale. Information-architecture or findability failures appear in 20.7\% of audited submissions, and cross-surface inconsistencies in 13.3\%. A patch can therefore be accurate in isolation and still fail within the larger documentation system.

\subsection{Trajectory analysis of failure root causes}
Patch-level labels describe what is wrong with the resulting documentation, but not why the agent produced it. We therefore inspected the trajectory behind each of the 1,267 audited submissions. For each material problem, we assigned one or more causes that the trace supports. Table~\ref{tab:trajectory-root-causes} reports submission-level rates across this full population.

\begin{table}[H]
\centering
\footnotesize
\setlength{\tabcolsep}{4pt}
\renewcommand{\arraystretch}{1.08}
\begin{tabular}{p{0.29\textwidth}p{0.49\textwidth}rr}
\toprule
\textbf{Trajectory-level root cause} & \textbf{Observable reasoning failure} &
\textbf{Submissions} & \textbf{Rate} \\
\midrule
Stopped at explaining the interface without examining how it is used in practice &
The agent described changed fields, settings, callbacks, or lifecycle mechanics, but did not test
the explanation against the reader's setup, decision, execution, verification, or recovery path. &
456 & 36.0\% \\

Did not inspect decisive evidence and filled the gap with a plausible assumption &
The agent found related material but stopped before the controlling implementation, schema, test,
or public contract, then completed the explanation with a convention that sounded reasonable. &
420 & 33.1\% \\

Stopped searching after finding the first plausible documentation surface &
The agent found a reasonable page to edit and did not continue checking other maintained,
generated, mirrored, migration, or workflow surfaces affected by the same change. &
382 & 30.1\% \\

Inspected relevant evidence but did not convert it into a complete coverage checklist &
The agent reached evidence bearing on the requirement but began drafting without tracking the
claims, setup, boundaries, examples, and reader actions that needed to survive into the final patch. &
349 & 27.5\% \\

Committed too early to a narrow interpretation of the task &
Before completing the investigation, the agent declared the task to be a rename, reference update,
single-page edit, or similarly narrow deliverable and ignored evidence outside that frame. &
346 & 27.3\% \\

Overgeneralized or misinterpreted partial evidence &
The agent inspected relevant evidence but converted one branch, example, implementation detail, or
deployment pattern into a broader or different public rule. &
333 & 26.3\% \\
\bottomrule
\end{tabular}
\captionsetup{font=footnotesize}
\caption{The six most common trajectory-level root causes across the 1,267 pre-rerun trajectories, which were frozen separately. Multiple causes may apply, so rates do not sum to 100\%. Table~\ref{tab:trajectory-root-causes-remaining} in the appendix lists the less common causes.}
\label{tab:trajectory-root-causes}
\end{table}

Agents often lack a reliable test for whether they have gathered enough evidence. Sometimes agents stop researching before they reach the decisive evidence. Other times they begin drafting from partial information without recognizing that their evidence is incomplete. This pattern suggests a failure to recognize uncertainty. Prior work reports similar findings~\citep{liu2025uncertainty,shao2025seekbench,liu2025cart}. Models struggle to identify the source of uncertainty. Information-seeking agents often answer before the available evidence is sufficient, and they do not reliably recognize when more information gathering has value.

Premature closure, shifting from investigation to drafting too soon, cuts across many of the root causes. Once an agent finds a plausible interpretation or a reasonable page to edit, it often starts drafting. It may stop before reaching key evidence, and it may also stop before checking every affected documentation surface, which leaves parts of the documentation stale. Insufficient search is only part of the problem. The larger part is that agents lack a reliable stopping rule. Such a rule would tell an agent when it understands the task, the evidence, and the documentation impact well enough to begin writing.

We found no clear relationship between the assigned root cause and trajectory length, whether measured by turns or by token use. Effort also did not rise with task difficulty. We defined an item's difficulty from the mergeability of the other six agents' patches on that item. Within each agent, the rank correlation between difficulty and effort was $+0.034$ for processed tokens, $+0.034$ for trace-event count, and $+0.030$ for tool actions. All task-clustered 95\% confidence intervals include zero. Premature closure therefore does not necessarily produce a short trajectory. An agent may stop investigating early and then spend substantial effort drafting, revising, or elaborating an incomplete account.

\section{Discussion}
\label{sec:discussion}

\subsection{Missing context about the reader}

Some failures that we attribute to a missing model of the reader may instead reflect missing context about how the software is used. Agents cannot always infer from parametric knowledge alone what readers are trying to accomplish or which details they need. That inference is especially hard when user motivations and the surrounding workflow context are implicit rather than stated.

\subsection{Coarse training rewards}

Premature-closure failures may be related to coarse reward signals during post-training. RAGEN finds that trajectory-level rewards do not reliably teach agents how to reason through multi-turn tasks. Without fine-grained, reasoning-aware feedback, agents may learn shallow strategies or produce reasoning that is not grounded in the environment~\citep{wang2025ragen}. Kim et al.\ report a similar pattern~\citep{kim2026soundreasoning}. In their experiments, outcome-only reinforcement learning improved final accuracy but made intermediate reasoning less accurate and less internally consistent. Models learned shortcuts rather than reliable reasoning procedures. These results offer possible explanations for our findings. The agents seemed to infer scope from early cues, such as the location of a code change or the name of a feature. They then began drafting within that narrow frame and filled evidence gaps with plausible assumptions that their environment did not support.

\subsection{Knowing when to stop investigating}

Another explanation is that deciding when the evidence is sufficient is itself a difficult capability. SeekBench reports that search agents trained with reinforcement learning answered before gathering sufficient evidence in 76.5\% of the evaluated trajectories~\citep{shao2025seekbench}. CaRT shows that models may rely on superficial stopping rules, such as the number of turns, instead of checking for a decisive fact~\citep{liu2025cart}. Related studies report that language models struggle to retract an earlier inference when new evidence contradicts it and tend to seek examples that confirm an initial hypothesis rather than examples that might disprove it~\citep{wilie2024beliefrevision,jhaveri2026falsify}. These findings match the patterns in our trajectory analysis.

\subsection{Additional guidance and scaffolding}
Several changes could plausibly address the observed failures: explicit instructions in the prompts that ask agents to consider the reader's goal, skills that emphasize task completion and findability, broader tools, scratch notes, and explicit verification. We explored these approaches informally but did not systematically compare them against a baseline, so we cannot conclude whether they improved documentation quality. Future work should test their effects through controlled comparisons.

\section{Limitations}
\label{sec:limitations}

\subsection{Measurement}
\label{sec:limitations-measurement}
Task-specific rubrics and LLM judges (Section~\ref{sec:scoring-model}) let us score open-ended documentation patches at scale. Because human validation covers only part of the evaluation, automated rubric generation and scoring may still introduce errors. We check commands and examples against the available evidence instead of running every documented procedure in its repository's native build and runtime environment. As a result, the benchmark has no deterministic checks for code samples and links.

We find that generated rubrics tend to contain more criteria than maintainer-authored rubrics. Many of these additional criteria identify valid documentation improvements, but maintainers may consider them less important. During human calibration, reviewers prioritized the noncritical P1--P3 criteria differently. Depending on repository norms, some emphasized style, while others placed less weight on it. Some preferred comprehensive documentation, whereas others favored a simple, easy-to-follow user path over broader coverage. These preferences do not support a single universal weighting of P1--P3 criteria. The score therefore weights all criteria equally in the mean and handles P0 failures separately through the score cap. The published dataset retains the P0--P3 labels, and the accompanying scoring code lets practitioners apply other weights.

Manual review found that some rubric criteria overlap and are not fully independent. Some overlap is warranted, because a single documentation failure can cause several related problems. In a later audit, we tried to merge overlapping criteria. Merging sometimes lost important distinctions, so we kept the overlapping criteria.

\subsection{No internet access}
\label{sec:limitations-internet}
We evaluated agents without internet access (Sections~\ref{sec:contamination} and~\ref{sec:baselines}). To assess how this restriction affected performance, we reviewed 683 trajectories from items on which no agent produced a mergeable patch. We found blocked network requests in 73 runs (10.7\%). In 63 of these runs, network access was not necessary to produce a correct patch. The requests mostly involved setup or validation, such as installing dependencies, building documentation, running formatters or linters, and parsing YAML or JSON configuration files. Only 10 runs (1.5\% of all audited trajectories) tried to retrieve external evidence that a correct patch required and the supplied inputs lacked.

Internet access might therefore have helped on a small number of items. However, in an earlier web-enabled pilot, 26 of 85 runs (30.6\%) retrieved the item's upstream pull request and its merged documentation. Given this contamination risk, we kept the reported evaluation sealed.

\subsection{Asymmetric label construction}
\label{sec:limitations-labels}
The evidence for the abstention and patch labels (Section~\ref{sec:dataset}) is not equally strong. Documentation-needed items often have direct evidence. Some abstention labels, by contrast, rely only on the absence of a related documentation change within a 90-day audit window. That absence does not necessarily show that documentation was unnecessary. An update may have been forgotten, or it may have happened after 90 days without a link to the code pull request. As a result, some items that needed documentation may be mislabeled as abstention items. This asymmetric label noise could distort the measured decision performance.

\subsection{Single-run evaluation}
\label{sec:limitations-single-run}
For budget and time reasons, we run each agent on each item once (Section~\ref{sec:baselines}). We therefore do not report \texttt{pass@k}, \texttt{pass\textasciicircum k}, best-of-$k$ performance, or within-item run-to-run variance. The results characterize one sampled trajectory per agent and item. They do not measure the probability that an agent reliably produces the same decision or documentation quality across repeated attempts.

\subsection{Low-information prose may be under-penalized}
\label{sec:limitations-ai-slop}
The rubric-based scoring (Section~\ref{sec:scoring-model}) and the patch-level failure rates (Section~\ref{sec:failure-modes}) may miss low-information prose. A general instruction to identify coherent but low-information prose flagged 3.2\% of submissions. A second prompt asked the judge to flag submissions with two or more specific patterns. The patterns included redundant paraphrases, unnecessary explanations, excessive bulleted lists, formulaic contrasts (not X, but Y), three-part constructions, and heavy use of em dashes. This prompt flagged 7.6\% of submissions. The increase suggests that LLM judges may miss low-information prose during scoring.

\section*{Disclosure}

Two authors are affiliated with Promptless, a company that builds documentation agents.

\section{Conclusion}

\bench\ shows that even frontier models paired with frontier coding harnesses cannot yet reliably produce expert-level user-facing documentation in one attempt. The highest composite score on the 117-item held-out split is 47.3 out of 100. Agents still misjudge whether documentation is needed, and when they do edit, their patches may be factually correct but miss what readers need. Fluent prose and capable repository tooling do not yet close this gap.

\bench\ measures this gap with 292 real items, drawn from software changes and reported documentation gaps, and it shows where decisions and patches fail. We release the evaluation harness, item schema, dataset card, and development examples so that others can build on the benchmark. Benchmark materials and release information are available at \url{https://dogbench.ai}.

\bibliographystyle{plainnat}


\appendix
\clearpage

\section{Dataset Construction and Task Examples}
\label{app:dataset-details}
\label{sec:pr-selection}
\label{sec:source-selection}
\label{sec:curation}
\label{app:worked-examples}
\label{app:schema}

\subsection{Additional dataset details}
For documentation-needed pull requests, we also require 5 to 500 changed lines of user-facing documentation across files. The lower bound is meant to exclude small, mechanical edits. The upper bound is meant to exclude broad rewrites and changes to the hosting framework.

Empirically, about six routine pull requests need no documentation update for every one that does. The benchmark's constructed class proportions (87 abstention items and 205 documentation-needed items) are therefore not a prevalence estimate.

\subsection{Task, patch, and evaluation case studies}
\label{app:case-studies}
\begingroup\fontsize{9}{10.5}\selectfont
\setlength{\LTpre}{10pt}\setlength{\LTpost}{10pt}
\setlength{\LTcapwidth}{\textwidth}
\captionsetup{font=small,justification=raggedright,singlelinecheck=false,skip=6pt}
\begin{longtable}{@{}p{\textwidth}@{}}
\caption{Pants: GLM 5.2+OpenCode documents coverage merging, and the saved evaluation passes all P0 criteria.}
\label{tab:case-pants}\\
\toprule\endfirsthead
\multicolumn{1}{l}{\small\textbf{Table \thetable\ continued: Pants coverage merging}}\\
\toprule\endhead\bottomrule\endfoot
\textbf{Setting}\\*
\textbf{Item:} \texttt{pantsbuild-pants-pr23219}.
\textbf{Agent:} GLM 5.2+OpenCode.
\textbf{Source:} \href{https://github.com/pantsbuild/pants/pull/23219}{Pants PR \#23219}.
This is a documentation-only item, so the agent receives no implementation code diff. An update is required.\\
\midrule
\textbf{Task request supplied to the agent}\\*
\begin{minipage}[t]{\linewidth}
\begin{lstlisting}[style=caseexcerpt]
Document combining Python coverage from sharded Pants test runs and applying the coverage threshold to the combined result.
\end{lstlisting}
\end{minipage}\\
\midrule
\textbf{Human merged patch (excerpts)}\\*
From \path{docs/docs/python/goals/test.mdx}. \texttt{[...]} marks omissions.\\*
\textit{Incomplete coverage, raw output, and threshold placement}\\*
\begin{minipage}[t]{\linewidth}
\begin{lstlisting}[style=caseexcerpt]
+When using the `--shard` flag to split tests across CI runners, each shard only exercises a fraction of your test targets. The per-shard coverage report will be artificially low. To get accurate coverage you need to combine the binary `.coverage` files from all shards.
[...]
+Add `"raw"` to your coverage reports so Pants writes the `.coverage` binary, and set `relative_files = true` so that `coverage combine` can match paths across different sandbox directories:
[...]
+:::caution Don't set `fail_under` in `[coverage-py]` when sharding
+Each shard only runs a fraction of your targets, so per-shard coverage is intentionally incomplete. Setting `fail_under` in `pants.toml` or `pants.ci.toml` will cause every shard to fail. Enforce the threshold after combining all shards instead.
+:::
\end{lstlisting}
\end{minipage}\\
\textit{Collection, combination, and reporting}\\*
\begin{minipage}[t]{\linewidth}
\begin{lstlisting}[style=caseexcerpt]
+After all shards complete, collect their `.coverage` binaries, combine them with `coverage combine`, and generate the final report. The following example uses GitHub Actions, but the same approach applies to any CI system:
[...]
+        coverage combine --rcfile=.coveragerc .coverage.shard*
+        coverage xml   --rcfile=.coveragerc -o coverage-report/coverage.xml
+        coverage report --rcfile=.coveragerc --fail-under=80
\end{lstlisting}
\end{minipage}\\
\textit{Per-shard \texttt{global\_report} caveat}\\*
\begin{minipage}[t]{\linewidth}
\begin{lstlisting}[style=caseexcerpt]
+:::note `global_report` and sharding
+With `[coverage-py] global_report = true`, per-shard reports show 0% for untouched files. Consider applying this setting only in the post-merge step rather than in `pants.ci.toml`.
+:::
\end{lstlisting}
\end{minipage}\\
\textit{Link from \texttt{advanced-target-selection.mdx}}\\*
\begin{minipage}[t]{\linewidth}
\begin{lstlisting}[style=caseexcerpt]
+When using `--shard` with test coverage enabled, each shard only exercises a fraction of your targets, producing artificially low coverage numbers. You need to combine the coverage data from all shards in a post-shard CI step to get accurate results. To learn how to do this for Python see [Coverage with test sharding](../python/goals/test.mdx#coverage-with-test-sharding) for the full configuration and CI workflow.
\end{lstlisting}
\end{minipage}\\
These excerpts omit the GitHub Actions upload and download steps and other configuration details. The judgments below assess the model-generated patch, not these human excerpts.\\
\midrule
\textbf{Model-generated patch (excerpts)}\\*
Each label names the rubric criterion below that the excerpt addresses. Excerpts come from \path{docs/docs/python/goals/test.mdx} unless noted.\\*
\textit{Why shard coverage must be combined (C1)}\\*
\begin{minipage}[t]{\linewidth}
\begin{lstlisting}[style=caseexcerpt,breakatwhitespace=false]
+Things are different when you shard your tests across multiple CI jobs with `--test-shard=k/N` (see [Sharding the input targets](../../using-pants/advanced-target-selection.mdx#sharding-the-input-targets)). Each shard is a separate `pants test` invocation, so each one generates its own coverage report and evaluates `fail_under` against only the tests in that shard. The per-shard numbers will usually be lower than the combined number, and a passing combined threshold can appear to fail.
+
+To apply `fail_under` to the coverage combined across _all_ shards, do the threshold check outside of Pants, on the union of the per-shard data:
\end{lstlisting}
\end{minipage}\\
\textit{Raw output and threshold placement (C3)}\\*
\begin{minipage}[t]{\linewidth}
\begin{lstlisting}[style=caseexcerpt]
+    [coverage-py]
+    report = ["raw"]
+    # Do not set fail_under here; apply it after combining instead.
\end{lstlisting}
\end{minipage}\\*
\textit{Preserving shard data, combining, and enforcing the threshold (C3)}\\*
\begin{minipage}[t]{\linewidth}
\begin{lstlisting}[style=caseexcerpt]
+    $ pants test --shard=0/3 --use-coverage ::
+    $ cp dist/coverage/python/.coverage .coverage.shard0
+    $ pants test --shard=1/3 --use-coverage ::
+    $ cp dist/coverage/python/.coverage .coverage.shard1
+    $ pants test --shard=2/3 --use-coverage ::
+    $ cp dist/coverage/python/.coverage .coverage.shard2
+    $ coverage combine .coverage.shard0 .coverage.shard1 .coverage.shard2
+    $ coverage report --fail-under=70
\end{lstlisting}
\end{minipage}\\
\textit{Link from the sharding guide (C10)}\\*
\begin{minipage}[t]{\linewidth}
\begin{lstlisting}[style=caseexcerpt]
+If you use test sharding together with Python coverage and a `[coverage-py].fail_under` threshold, note that each shard evaluates the threshold against its own coverage only. See [Combining coverage from sharded test runs](../python/goals/test.mdx#combining-coverage-from-sharded-test-runs) in the Python test docs for how to apply the threshold to the coverage combined across all shards.
\end{lstlisting}
\end{minipage}\\
The human patch supplies a GitHub Actions workflow, while the model supplies a generic numbered procedure and shell commands. The rubric does not require matching the human implementation. The full model patch omits the \texttt{global\_report} caveat (C15).\\
\midrule
\begin{minipage}[t]{\linewidth}
\textbf{Selected rubric criteria and verdicts}\par\smallskip Full criteria, including pass and fail conditions, are available on the\href{https://dogbench.ai}{benchmark website}.\par\smallskip
\renewcommand{\arraystretch}{1.08}
\begin{tabular}{@{}p{0.11\linewidth}p{0.10\linewidth}p{0.72\linewidth}@{}}
\textbf{ID / priority} & \textbf{Verdict} & \textbf{Rubric requirement}\\
C1 / P0 & Pass & The patch must explicitly document that separate Python \texttt{pants test --shard=k/N} invocations each measure only their shard’s tests, so an individual shard’s coverage is incomplete, and accurate project coverage requires combining data from every shard before producing the final report.\\[3pt]
C3 / P0 & Pass & After reading, a user must know to: 1. Preserve the \texttt{.coverage} data from every shard. 2. Collect and combine all shard data in a post-shard job. 3. Generate the final report only after combination and enforce its threshold on the combined result, not on each incomplete shard through per-shard \texttt{[coverage-py].fail\_under} configuration.\\[3pt]
C10 / P1 & Pass & The maintained sharding guidance must include a nearby, discoverable connection to the detailed Python coverage-and-sharding guidance. At the pinned base, the natural source is \texttt{docs/docs/using-pants/advanced-target-selection.mdx} near \texttt{\#\# Sharding the input targets}; an equivalent maintained successor surface passes. The connection must resolve to the actual maintained detailed section. Any repository-supported link form, route, or anchor that resolves correctly passes. A relative \texttt{.mdx} link such as \texttt{../python/goals/test.mdx\#coverage-with-test-sharding} is the base-tree conventional example, not the only acceptable spelling.\\[3pt]
C15 / P2 & Fail & The patch must explain that per-shard \texttt{[coverage-py].global\_report = true} can report 0\% for files untouched by that shard and should not be treated as final project coverage.\\[3pt]
\end{tabular}\end{minipage}\\
\textbf{Full-patch outcome:} correct patch decision; \textbf{92.3/100}; \textbf{P0-clean}.\\
\end{longtable}\endgroup

\begingroup\fontsize{9}{10.5}\selectfont
\setlength{\LTpre}{10pt}\setlength{\LTpost}{10pt}
\setlength{\LTcapwidth}{\textwidth}
\captionsetup{font=small,justification=raggedright,singlelinecheck=false,skip=6pt}
\begin{longtable}{@{}p{\textwidth}@{}}
\caption{Jujutsu: GPT-5.6 Sol+Codex adds the new type but omits conversion semantics and an affected return type.}
\label{tab:case-jj}\\
\toprule\endfirsthead
\multicolumn{1}{l}{\small\textbf{Table \thetable\ continued: Jujutsu byte strings}}\\
\toprule\endhead\bottomrule\endfoot
\textbf{Setting}\\*
\textbf{Item:} \texttt{jj-vcs-jj-pr9347}.
\textbf{Agent:} GPT-5.6 Sol+Codex.
\textbf{Source:} \href{https://github.com/jj-vcs/jj/pull/9347}{Jujutsu PR \#9347}.
In this code-triggered item, the change introduces a byte-string template type. The agent receives the code diff and the pre-change documentation and must update \path{docs/templates.md}.\\
\midrule
\textbf{Agent-visible code diff (excerpts)}\\*
\textit{Annotation-line content: \texttt{cli/src/commit\_templater.rs}}\\*
\begin{minipage}[t]{\linewidth}
\begin{lstlisting}[style=caseexcerpt,breakatwhitespace=false]
             let out_property = self_property.map(|line| line.content);
-            // TODO: Add Bytes or BString template type?
-            Ok(P::wrap_template(out_property.into_template()))
+            Ok(out_property.into_dyn_wrapped())
\end{lstlisting}
\end{minipage}\\
\textit{Fallible byte-to-string conversion: \texttt{cli/src/template\_builder.rs}}\\*
\begin{minipage}[t]{\linewidth}
\begin{lstlisting}[style=caseexcerpt,breakatwhitespace=false]
+        let from_bytes =
+            |s: BString| Ok(String::from_utf8(s.into()).map_err(|err| err.utf8_error())?);
+        let property = match self.property.try_into_string() {
+            Ok(string_property) => return Some(string_property),
+            Err(property) => property,
+        };
+        let property = match property.try_into_byte_string() {
+            Ok(bytes_property) => return Some(bytes_property.and_then(from_bytes).into_dyn()),
+            Err(property) => property,
+        };
\end{lstlisting}
\end{minipage}\\
The annotation method stops wrapping its result as a \texttt{Template}.
The conversion path uses \texttt{String::from\_utf8} and propagates invalid-UTF-8
errors. These changes motivate the return-type and conversion documentation.\\
\midrule
\textbf{Human merged patch (excerpts from \texttt{docs/templates.md})}\\*
\textit{Annotation-line return type}\\*
\begin{minipage}[t]{\linewidth}
\begin{lstlisting}[style=caseexcerpt,breakatwhitespace=false]
-* `.content() -> Template`: Line content including newline character.
+* `.content() -> ByteString`: Line content including newline character.
\end{lstlisting}
\end{minipage}\\
\textit{Conversion to byte strings}\\*
\begin{minipage}[t]{\linewidth}
\begin{lstlisting}[style=caseexcerpt,breakatwhitespace=false]
+### `ByteStringify` type
+
+An expression that can be converted to a `ByteString`.
+
+A `String` can be converted to a `ByteString` losslessly. Any types that can be
+converted to `Template` can also be `ByteStringify`. Unlike `Template`, color
+labels are stripped.
\end{lstlisting}
\end{minipage}\\
\textit{Conversion to UTF-8 strings}\\*
\begin{minipage}[t]{\linewidth}
\begin{lstlisting}[style=caseexcerpt,breakatwhitespace=false]
 Any types that can be converted to `Template` can also be `Stringify`. Unlike
-`Template`, color labels are stripped.
+`Template`, color labels are stripped. Invalid UTF-8 sequences are rejected.
\end{lstlisting}
\end{minipage}\\
The rest of the human patch introduces \texttt{ByteString} and its methods, updates regex-capture return types, and revises the equality documentation.\\
\midrule
\textbf{Model-generated patch (excerpts)}\\*
Each label names the rubric criterion below that the excerpt addresses.\\*
\textit{New type in the maintained reference (C1, C4)}\\*
\begin{minipage}[t]{\linewidth}
\begin{lstlisting}[style=caseexcerpt,breakatwhitespace=false]
+### `ByteString` type
+
+_Conversion: `Boolean`: yes, `Serialize`: yes, `Template`: yes_
+
+A byte string can contain arbitrary bytes and does not have to be valid UTF-8.
+It can be implicitly converted to `Boolean`. When converted to a template, its
+bytes are preserved without a UTF-8 conversion. The following methods are
+defined.
\end{lstlisting}
\end{minipage}\\
\textit{Conversion definition (C7)}\\*
\begin{minipage}[t]{\linewidth}
\begin{lstlisting}[style=caseexcerpt,breakatwhitespace=false]
+### `ByteStringify` type
+
+An expression that can be converted to a `ByteString`. This includes
+`ByteString` values and any type that can be converted to `Stringify`.
\end{lstlisting}
\end{minipage}\\
\textit{Updated regex-capture signatures (C9)}\\*
\begin{minipage}[t]{\linewidth}
\begin{lstlisting}[style=caseexcerpt,breakatwhitespace=false]
-* `.get(index: Integer) -> String`: Returns the capture group at `index`.
+* `.get(index: Integer) -> ByteString`: Returns the capture group at `index`.
   Capture group 0 is the full match. Errors if the index is out of bounds.
-* `.name(name: Stringify) -> String`: Returns the named capture group `name`.
+* `.name(name: Stringify) -> ByteString`: Returns the named capture group `name`.
\end{lstlisting}
\end{minipage}\\
The model describes bytes that may not be valid UTF-8 and passes the revised C4. Its conversion definition does not explain formatted template output or label stripping (C7, revised to P1). It updates the two regex-capture signatures but leaves \texttt{AnnotationLine.content()} unchanged (C9). It also leaves the \texttt{Stringify} section unchanged, so it omits invalid-UTF-8 rejection (C6). The full patch has these omissions, not only the excerpts.\\
\midrule
\begin{minipage}[t]{\linewidth}
\textbf{Selected rubric criteria and verdicts}\par\smallskip
We revised C4, C6, and C7 to separate the type definition, UTF-8 rejection, and conversion to bytes. This case reports an evaluation under the revised criteria. Full criteria and pass and fail conditions are available on the \href{https://dogbench.ai}{benchmark website}.\par\smallskip
\renewcommand{\arraystretch}{1.08}
\begin{tabular}{@{}p{0.11\linewidth}p{0.10\linewidth}p{0.72\linewidth}@{}}
\textbf{ID / priority} & \textbf{Verdict} & \textbf{Rubric requirement}\\
C1 / P0 & Pass & \texttt{docs/templates.md} is updated as the primary reference surface and introduces \texttt{ByteString} as a public template-language type.\\[3pt]
C4 / P0 & Pass & Explain that \texttt{ByteString} can contain bytes that are not valid UTF-8. A concise statement such as "arbitrary bytes" or "not guaranteed to be valid UTF-8" is sufficient; the phrase "ASCII-compatible" and an explicit comparison sentence with \texttt{String} are not required.\\[3pt]
C6 / P0 & Fail & Explain that converting a \texttt{ByteString} through \texttt{Stringify}/\texttt{stringify()} requires valid UTF-8 and rejects invalid byte sequences. This is a conversion precondition, not another test of the byte-string definition. A short warning, an error example, or a clear exception to the existing Template-to-Stringify statement is sufficient.\\[3pt]
C7 / P1 & Fail & Define \texttt{ByteStringify} as accepting values convertible to bytes and explain that strings and formatted template output can supply those bytes without retaining formatting labels. Equivalent descriptions or examples count; the literal word "lossless" is not required. A precise cross-reference to existing conversion documentation can supply these facts. This criterion concerns conversion to bytes, not rejection when converting bytes to UTF-8 strings (C6).\\[3pt]
C9 / P0 & Fail & The discoverable API reference in \texttt{docs/templates.md} shows all three signatures: \texttt{AnnotationLine.content() -\textgreater{} ByteString}, \texttt{RegexCaptures.get(index: Integer) -\textgreater{} ByteString}, and \texttt{RegexCaptures.name(name: Stringify) -\textgreater{} ByteString}. If the patch edits the global \texttt{replace(pattern, content, replacement)} documentation, it preserves that function's signature.\\[3pt]
\end{tabular}\end{minipage}\\
\textbf{Full-patch outcome:} correct patch decision; \textbf{44.4/100};
\textbf{P0 failures C6, C9}.\\
\end{longtable}\endgroup

\section{Rubric Construction and Human Validation}
\label{app:evaluation-details}
\label{sec:generated-rubric}
\label{app:verifiers}
\label{sec:scoring-gate}

\subsection{Research, audit, and debate}
Before adopting rubric-based evaluation, we explored two other approaches. First, we generated questions from the source pull request and checked whether the candidate documentation let a question-answering agent answer them. The questions were often too broad, covering documentation beyond the evaluated patch, or limited to technical details that did not reflect readers' goals. Second, we translated documentation into executable tests. This approach was particularly useful for procedural content, but test outcomes were hard to attribute to the patch under evaluation. A test could fail because of unchanged documentation, environment issues, or assumptions that the test generator introduced. We arrived at the current design through experiments, ablations, and input from human experts.

\paragraph{Research and synthesis}
Rubric construction begins with a research agent (GPT-5.6 Sol) that has internet, shell, and browser access. The research agent can install software, run examples, and test behavior to understand the reader's experience. The human-authored patch is not included among the supplied candidate patches, but internet research may uncover it. Every proposed criterion must have independent supporting evidence; a criterion justified only by the human patch is not allowed. The research agent produces a report that covers software behavior, reader needs, affected documentation, dependencies, and relevant constraints.

\paragraph{Audit and debate}
During the audit, each disputed criterion needs specific evidence, such as a source-file location, a test, or a maintainer comment. A criterion without that evidence is withdrawn. Each proposed change must also explain its consequence for the reader. This process can uncover errors in the underlying research or rubric synthesis, as the following example shows.

\paragraph{Example: distinguishing useful detail from required coverage}
Doc Detective PR~\#146 lets users pass multiple input paths in one comma-separated \texttt{--input} argument. Criterion C5 concerned whether the documentation should also explain that URL inputs are preserved rather than resolved as local paths (``URL passthrough''). The dispute concerned C5's priority. Both sides agreed that the behavior exists. The excerpts below come from the saved debate, with omissions marked \texttt{[...]}.

\par\medskip
\begingroup
\setlength{\fboxsep}{10pt}
\noindent\fcolorbox{black!25}{black!2}{%
\begin{minipage}{\dimexpr\linewidth-2\fboxsep-2\fboxrule\relax}
\small\raggedright
\textbf{Auditor, opening audit: lower C5 to P3}\par\smallskip
\textbf{Reasoning:} URL passthrough is real, but the criterion overstates its importance. The PR body and issue frame the feature as multiple CLI input paths; the tests added for this PR cover local file inputs, not URLs. [...] Revise this from a required P2 criterion into a P3 credit-only quality signal for patches that accurately mention URL values.
\end{minipage}%
}
\endgroup
\par\medskip

\par\medskip
\begingroup
\setlength{\fboxsep}{10pt}
\noindent\fcolorbox{black!25}{black!2}{%
\begin{minipage}{\dimexpr\linewidth-2\fboxsep-2\fboxrule\relax}
\small\raggedright
\textbf{Author, first response: retain P2}\par\smallskip
\textbf{C5 — REBUT, but partially concede.} [...] When the majority of real candidate patches independently document a behavior and omitting it is judged undesirable, that's evidence it's a genuine "good-patch" signal, not P3 excellence trivia.

[...] Given the rules say P2 = "a good patch includes it but failing it is still shippable" — that is \emph{exactly} the correct semantics for URL passthrough per both my report and the differential findings. P3 would understate it. [...]
\end{minipage}%
}
\endgroup
\par\medskip

\par\medskip
\begingroup
\setlength{\fboxsep}{10pt}
\noindent\fcolorbox{black!25}{black!2}{%
\begin{minipage}{\dimexpr\linewidth-2\fboxsep-2\fboxrule\relax}
\small\raggedright
\textbf{Auditor, second audit: task evidence does not justify P2}\par\smallskip
I still rebut C5 at P2. [...] The author’s differential evidence can justify it as desirable, but not enough to make it a P2 “good patch includes this” criterion when the issue, PR body, README context, and tests all center local/path multi-input syntax. Also, the current scoring already says a “Good” patch may have C5 partial or failed, which is internally inconsistent with keeping C5 as a required P2.

[...]
\end{minipage}%
}
\endgroup
\par\medskip

\par\medskip
\begingroup
\setlength{\fboxsep}{10pt}
\noindent\fcolorbox{black!25}{black!2}{%
\begin{minipage}{\dimexpr\linewidth-2\fboxsep-2\fboxrule\relax}
\small\raggedright
\textbf{Author, second response: concede P3}\par\smallskip
CONCEDE on C5. [...] Demoting to P3 keeps the credit the differential evidence supports while resolving the internal inconsistency the auditor identified in the scoring guide.
\end{minipage}%
}
\endgroup
\par\medskip

\subsection{Human validation for the rubrics and scorer}
Project maintainers first submitted free-form, GitHub-style inline review comments on the candidate patches without knowing which patch was which. Claude Sonnet 4.6 used those comments and the lines they annotated to draft structured rubrics. Each maintainer then edited, added, or deleted criteria, approved each criterion individually, and decided whether each candidate patch passed or failed each criterion. This process was separate from automatic rubric generation. Maintainers never saw the corresponding generated rubrics, and the rubric agents never saw the maintainer rubrics. One repository maintainer reviewed each item.

We did not rely only on maintainer rubrics, for two reasons. First, the cost and time were prohibitive for the scope of this study. Constructing one maintainer rubric took roughly 30 minutes to an hour, even for maintainers familiar with the domain. Second, expert review often did not give a consistent or exhaustive standard across items. Maintainers had different editorial priorities. Some emphasized style, and another emphasized information architecture. One favored self-contained pages, while another preferred progressive disclosure. Each is a defensible choice, but with only one maintainer per repository, those preferences change what the benchmark measures. Recruiting several maintainers for every repository was impractical. Maintainer rubrics can also contain omissions or mistakes, especially when reviewers must anticipate gaps that no candidate patch addresses.

We therefore generated task-specific rubrics from task-specific research and a consistent set of expert-reviewed technical-writing principles. We then validated the generated rubrics against the maintainer rubrics. This approach made evaluation at scale possible while keeping expert judgment central to its design and validation.

The validation set contains 42 maintainer-reviewed decisions: 23 items require a documentation change and 19 require no change. The 23 documentation-needed items cover 242 maintainer-validated criteria. Table~\ref{tab:doc-gate-validation} compares the automatic documentation-need gate with the maintainer decisions.

\begin{table}[H]
\centering
\begin{tabular}{lrrrrrrrrrr}
\toprule
\textbf{Subset} & \textbf{$n$} & \textbf{Acc.} & \textbf{Prec.} & \textbf{Recall} & \textbf{F1} & \textbf{TP} & \textbf{FP} & \textbf{FN} & \textbf{TN} \\
\midrule
Documentation-need gate & 42 & 0.976 & 0.958 & 1.000 & 0.979 & 23 & 1 & 0 & 18\\
\bottomrule
\end{tabular}
\captionsetup{font=footnotesize}
\caption{Agreement between the documentation-need gate and maintainer decisions on the 42 maintainer-reviewed items. Every item has a prediction.}
\label{tab:doc-gate-validation}
\end{table}

\begin{table}[H]
\centering
\begin{tabular}{lrr}
\toprule
\textbf{Human-rubric recall} & \textbf{Count} & \textbf{Percentage} \\
\midrule
Covered & 220 & 90.9\% \\
Missing & 17 & 7.0\% \\
Conflicting & 5 & 2.1\% \\
\midrule
Total & 242 & 100.0\% \\
\bottomrule
\end{tabular}
\captionsetup{font=footnotesize}
\caption{Coverage of 242 maintainer-validated criteria across 23 documentation-change examples.}
\label{tab:rubric-validation}
\end{table}

The generated rubrics cover 90.9\% of 242 maintainer-validated criteria.
In the reverse comparison, 59.8\% of 361 generated criteria are covered in
the maintainer-reference rubrics. Of the 137 criteria absent from the human rubrics, 76 are
routine or defensive checks that reviewers often leave implicit. These include prose and
markup conventions, repository conventions, and safeguards against fabricated content. Lack
of a match therefore does not establish that a generated criterion is invalid. After the
human-authored rubrics were complete, experts reviewed the generated rubrics and often agreed
with additional criteria they had not identified initially. These follow-up reviews were
qualitative and limited in scale.
\begin{table}[H]
\centering
\begin{tabular}{lrr}
\toprule
\textbf{Generated-rubrics precision} & \textbf{Count} & \textbf{Percentage} \\
\midrule
Covered & 216 & 59.8\% \\
Missing & 137 & 38.0\% \\
Conflicting & 8 & 2.2\% \\
\midrule
Total & 361 & 100.0\% \\
\bottomrule
\end{tabular}
\captionsetup{font=footnotesize}
\caption{Overlap precision of the generated rubrics against the maintainer rubrics. Full and partial nonconflicting matches are combined. A partial match does not validate every requirement in a generated criterion. Absence from the maintainer rubric does not show that a criterion is invalid. All conflicts are partial.}
\label{tab:generated-rubric-precision}
\end{table}

We also applied the generated and maintainer rubrics to the same candidate patches. The two sets of scores have a pooled Spearman correlation of 0.755 and an interval Krippendorff $\alpha$ of 0.805. They agree on 85.0\% of within-item candidate orderings. These measures suggest that evaluations based on the two kinds of rubric agree.

\subsection{Direct criterion-verdict validation}
\label{app:scorer-validation}

The scorer-validation study covers 20 documentation-needed items and 330 rubric criteria. For each item, GPT-5.6 Terra, Claude Sonnet 5, and a human reviewer judged the same anonymous candidate patch against the same rubric criteria. Requirements and triggered conditional criteria receive Pass or Fail. Deduction-only guardrails are marked violated or not violated. Table~\ref{tab:grader-human-validation} reports agreement with the human reference.

\begin{table}[H]
\centering
\begin{tabular}{lrrr}
\toprule
\textbf{Grader} & \textbf{Coverage} & \textbf{Matches} & \textbf{Agreement} \\
\midrule
GPT-5.6 Terra    & 20/20 & 310/330 & 93.9\% \\
Claude Sonnet 5 & 20/20 & 301/330 & 91.2\% \\
\bottomrule
\end{tabular}
\captionsetup{font=footnotesize}
\caption{Criterion agreement between each grader and the human reference after adjudication, across 20 validation items.}
\label{tab:grader-human-validation}
\end{table}

\subsection{Remaining P0 failures on human merged patches}

Table~\ref{tab:human-reference-failure-examples} shows four human merged patches that fail a P0 criterion because of a concrete defect.

\begingroup
\small
\setlength{\tabcolsep}{4pt}
\renewcommand{\arraystretch}{1.15}
\begin{longtable}{@{}p{0.18\textwidth}p{0.34\textwidth}p{0.43\textwidth}@{}}
\caption{Four examples of human merged patches with concrete P0 failures. The first column gives the criterion
ID and the merged patch's rubric score out of 100.}
\label{tab:human-reference-failure-examples}\\
\toprule
\textbf{Case / criterion / score} & \textbf{Faulty human-patch excerpt} & \textbf{Why the failure remains P0} \\
\midrule
\endfirsthead
\toprule
\textbf{Case / criterion / score} & \textbf{Faulty human-patch excerpt} & \textbf{Why the failure remains P0} \\
\midrule
\endhead
\bottomrule
\endfoot
\href{https://github.com/pantsbuild/pants/pull/22034}{Pants \#22034}\newline C8; 50.0
& \raggedright \texttt{pants experimental-deploy}\newline
  \texttt{src/k8s/:webpages}
& The exact-version address parser rejects the empty path component before the
colon, so the deployment command cannot resolve its target. Removing the slash
fixes it: \texttt{src/k8s:webpages}. The defect is small but blocks the advertised
operation. \\
\midrule
\href{https://github.com/opencost/opencost-website/pull/102}{OpenCost \#102}\newline C6; 44.4
& \raggedright Secret creation: \texttt{kubectl create secret generic azure-service-key -n kubecost}
(excerpt).\newline
Workload update: \texttt{helm upgrade opencost . --namespace opencost -f values.yaml}
& The Secret is created in \texttt{kubecost}, but the workload that mounts it is
updated in \texttt{opencost}. A workload cannot mount a Secret from another
namespace. The supplied credential-injection procedure therefore fails unless
the namespaces are made consistent. \\
\midrule
\href{https://github.com/strawberry-graphql/strawberry/pull/4342}{Strawberry \#4342}\newline C6; 31.2
& \raggedright Resolver: \texttt{def create\_user(self, email: str) -> str:}\newline
\texttt{\phantom{xx}return email}\newline
Schema: \texttt{strawberry.Schema(}\newline
\texttt{mutation=Mutation,}\newline
\texttt{extensions=[}\newline
\texttt{PydanticErrorExtension()}\newline
\texttt{],)}
& The usage example omits the required query root and never invokes Pydantic
validation. It cannot produce the advertised validation errors. The same PR's
working test supplies both a query root and Pydantic model construction,
which gives a direct implementation contrast. \\
\midrule
\href{https://github.com/dlt-hub/dlt/pull/2292}{dlt \#2292}\newline C8; 55.6
& \raggedright \texttt{iceberg\_tables[}\newline
\texttt{"my\_iceberg\_table"]}\newline
\texttt{.optimize.compact()}
& The helper returns native PyIceberg Table objects. The retained source check
for supported version 0.8.1 finds no \texttt{optimize} API or dynamic fallback.
The copied Delta-style operation cannot run on that object, so a supported Iceberg
operation must replace it. \\
\end{longtable}
\endgroup

\section{Additional Results and Robustness Analyses}
\label{app:analysis-details}
\label{sec:ablations}

\subsection{Population-specific robustness results}

The paper's primary comparison uses the 117-item held-out split. Table~\ref{tab:full-robustness} reports all seven agent lanes on the full 292-item population.

\begin{table}[H]
\centering
\small
\setlength{\tabcolsep}{4.5pt}
\resizebox{\textwidth}{!}{%
\begin{tabular}{lrrrrrrr}
\toprule
\textbf{Agent} & \textbf{Score} & \textbf{Accuracy} &
\textbf{Patch recall} & \textbf{Abstention recall} & \textbf{P0-clean delivery} & \textbf{Delivered quality} &
\textbf{Conditional quality} \\
\midrule
Qwen3.8 Max+OpenCode & 48.1 & 81.2 & 84.4 & 73.6 & 30.7 & 35.8 & 42.4 \\
GLM 5.2+OpenCode & 46.9 & 84.6 & 84.9 & 83.9 & 26.3 & 32.6 & 38.4 \\
GPT-5.6 Sol+Codex & 45.0 & 81.2 & 96.1 & 46.0 & 38.5 & 44.1 & 45.9 \\
Claude Opus 4.8+Claude Code & 42.2 & 80.8 & 79.0 & 85.1 & 21.0 & 28.1 & 35.5 \\
Kimi K2.7 Code+OpenCode & 42.0 & 79.1 & 88.8 & 56.3 & 25.9 & 33.5 & 37.7 \\
Claude Sonnet 4.6+Claude Code & 39.1 & 81.5 & 94.6 & 50.6 & 25.9 & 31.8 & 33.6 \\
GPT-5.5+Codex & 32.3 & 75.0 & 95.6 & 26.4 & 35.6 & 41.6 & 43.5 \\
\bottomrule
\end{tabular}
}
\captionsetup{font=footnotesize}
\caption{Full-population results for seven agent lanes on all 292 items (205 documentation-needed and 87 abstention items), in percent. Definitions match Table~\ref{tab:headline-composite}. Rows are ordered by the unrounded composite score.}
\label{tab:full-robustness}
\end{table}

\paragraph{Patch or abstention decision errors}
Table~\ref{tab:decision-confusion} counts decision errors on all 292 items, with \texttt{patch} as the positive class. A false negative (FN) is an abstention on a documentation-needed item. A false positive (FP) is a patch on an abstention item.

\begin{table}[H]
\centering
\small
\begin{tabular}{lrr}
\toprule
\textbf{Agent} & \textbf{FN} & \textbf{FP} \\
\midrule
GLM 5.2+OpenCode & 31 & 14 \\
Claude Sonnet 4.6+Claude Code & 11 & 43 \\
Qwen3.8 Max+OpenCode & 32 & 23 \\
GPT-5.6 Sol+Codex & 8 & 47 \\
Claude Opus 4.8+Claude Code & 43 & 13 \\
Kimi K2.7 Code+OpenCode & 23 & 38 \\
GPT-5.5+Codex & 9 & 64 \\
\bottomrule
\end{tabular}
\caption{Decision error counts for seven agent lanes on the full 292-item population.}
\label{tab:decision-confusion}
\end{table}

\paragraph{Repository-footprint ablation}
We measured repository popularity from GitHub on August 18, 2026. The comparison uses all 205 documentation-needed items and the same 82 low-footprint and 91 popular items for every agent (Table~\ref{tab:popularity-ablation}). It excludes the 32 items from repositories with 1,000 to 4,999 stars.

\begin{table}[H]
\centering
\small
\begin{tabular}{lrrrr}
\toprule
\textbf{Agent} & \textbf{Low-footprint} & \textbf{Popular} & \textbf{$\Delta$ low$-$popular} & \textbf{Repo 95\% CI} \\
\midrule
Claude Opus 4.8+Claude Code & 26.3 & 28.5 & $-2.2$ & [$-8.1$, 5.9] \\
Claude Sonnet 4.6+Claude Code & 31.7 & 29.9 & $+1.8$ & [$-3.3$, 7.2] \\
GLM 5.2+OpenCode & 33.6 & 29.9 & $+3.6$ & [$-4.4$, 12.9] \\
GPT-5.5+Codex & 45.7 & 36.0 & $+9.7$ & [$+0.8$, 16.1] \\
GPT-5.6 Sol+Codex & 47.2 & 42.4 & $+4.8$ & [$-3.4$, 10.9] \\
Kimi K2.7 Code+OpenCode & 34.2 & 32.5 & $+1.8$ & [$-5.9$, 9.3] \\
Qwen3.8 Max+OpenCode & 35.8 & 34.2 & $+1.7$ & [$-6.6$, 9.2] \\
\bottomrule
\end{tabular}
\caption{Mean documentation-quality score by repository-popularity band for 173 of the 205
documentation-needed items. Low-footprint repositories have fewer than 1,000 stars, and popular repositories have at least 5,000. Intervals cover the low-minus-popular difference in means from 20,000 percentile bootstrap resamples of repositories within each band.}
\label{tab:popularity-ablation}
\end{table}

\section{Failure Taxonomy: Additional Categories}
\label{app:failure-taxonomy}

\subsection{Additional artifact-level failure categories}
Table~\ref{tab:full-patch-failure-taxonomy} lists the artifact-level failure categories that Table~\ref{tab:patch-failure-modes} omits.

\begingroup
\footnotesize
\setlength{\tabcolsep}{5pt}
\renewcommand{\arraystretch}{1.08}
\begin{longtable}{p{0.25\textwidth}p{0.55\textwidth}r}
\caption{Additional artifact-level categories, continuing Table~\ref{tab:patch-failure-modes}, with patch-level rates across the 1,267 submissions audited before the reruns. Multiple categories may apply.}
\label{tab:full-patch-failure-taxonomy}\\
\toprule
\textbf{Technical-writing failure mode} & \textbf{Effect on the documentation} & \textbf{Rate} \\
\midrule
\endfirsthead
\toprule
\textbf{Technical-writing failure mode} & \textbf{Effect on the documentation} & \textbf{Rate} \\
\midrule
\endhead
\bottomrule
\endfoot
Low information density or poor scannability & Repeats information, adds unnecessary
structure, or uses disproportionately long prose for the information conveyed. & 7.6\% \\
Fabricated content & Invents an interface, command, control, behavior, version requirement,
or guarantee; distinct from a false description of a real interface. & 6.1\% \\
Nonfunctional example & Supplies a code block, command, configuration, or API example that
would fail or teach the wrong call shape. & 2.5\% \\
Documentation-system defect & Breaks links, markup, rendering, terminology, or documentation-system
conventions. & 0.9\% \\
Ambiguous or contradictory guidance & Gives incompatible instructions or leaves a material
rule ambiguous within the edited documentation. & 0.4\% \\
Scope creep & Edits unrelated files or topics beyond the documentation need; excludes
companion edits required for consistency. & 0.2\% \\
\end{longtable}
\endgroup

\subsection{Less common trajectory-level causes}
Table~\ref{tab:trajectory-root-causes-remaining} lists the trajectory-level causes that Table~\ref{tab:trajectory-root-causes} omits.

\begin{table}[H]
\centering
\footnotesize
\setlength{\tabcolsep}{3pt}
\renewcommand{\arraystretch}{1.08}
\caption{Less common trajectory-level root causes, continuing Table~\ref{tab:trajectory-root-causes}, across the 1,267 pre-rerun trajectories, which were frozen separately.}
\label{tab:trajectory-root-causes-remaining}
\begin{tabular}{p{0.31\textwidth}p{0.49\textwidth}rr}
\toprule
\textbf{Trajectory-level root cause} & \textbf{Observable reasoning failure} &
\textbf{Submissions} & \textbf{Rate} \\
\midrule
Found the correct fact, then dropped or contradicted it during drafting &
The correct distinction appeared in the evidence or reasoning but disappeared, weakened, or
reversed in the patch. & 81 & 6.4\% \\
Selected the wrong or conflicting source as authoritative &
The agent trusted stale documentation, generated output, an internal representation, or permissive
runtime behavior over the maintained public contract. & 72 & 5.7\% \\
Validated presentation or file mechanics, but not the substantive claim &
The agent checked syntax, links, formatting, or file existence without validating the underlying
command, example, route, or factual statement. & 50 & 3.9\% \\
Did not perform a reader-priority and compression pass &
The agent stopped after inserting relevant content without removing repetition, artificial
structure, or low-density prose. & 22 & 1.7\% \\
Did not run documentation-specific validation &
The agent omitted the relevant link, markup, navigation, generation, spelling, or build check. &
9 & 0.7\% \\
Lost requirements or edit state during a long trajectory &
Earlier requirements vanished after extended investigation, rewriting, or scope expansion. &
7 & 0.6\% \\
Did not recover relevant historical context &
The agent missed a relevant issue, pull request, release milestone, comment, or design decision. &
5 & 0.4\% \\
Did not maintain a boundary between required work and unrelated expansion &
The agent expanded into adjacent topics or files without tying each edit to the reader need. &
2 & 0.2\% \\
\bottomrule
\end{tabular}
\end{table}


\end{document}